\documentclass[sigconf,natbib=true]{acmart}
\usepackage{multirow}

\usepackage{graphicx}
\usepackage{textcomp}
\usepackage{xcolor}
\usepackage{multirow} 
\usepackage{multicol} 
\usepackage{algorithm}
\usepackage{algpseudocode}

\AtBeginDocument{%
  \providecommand\BibTeX{{%
    \normalfont B\kern-0.5em{\scshape i\kern-0.25em b}\kern-0.8em\TeX}}}

\copyrightyear{2024}
\acmYear{2024}
\setcopyright{acmlicensed}
\acmConference[WWW '24] {Proceedings of the ACM Web Conference 2024}{May 13--17, 2024}{Singapore, Singapore.}
\acmBooktitle{Proceedings of the ACM Web Conference 2024 (WWW '24), May 13--17, 2024, Singapore, Singapore}
\acmPrice{15.00}
\acmISBN{979-8-4007-0171-9/24/05}
\acmDOI{10.1145/XXXXXX.XXXXXX}

\acmConference[Conference acronym 'XX]{Make sure to enter the correct
  conference title from your rights confirmation emai}{June 03--05,
  2018}{Woodstock, NY}
\acmPrice{15.00}
\acmISBN{978-1-4503-XXXX-X/18/06}

\begin{document}
\fancyhead{}
\title{Towards Personalized Privacy: User-Governed Data Contribution for Federated Recommendation}

\author{Liang Qu}
\email{liang.qu@uq.edu.au}
\affiliation{%
  \institution{The University of Queensland}
  \city{Brisbane}
  \country{Australia}
}

\author{Wei Yuan}
\email{w.yuan@uq.edu.au}
\affiliation{%
  \institution{The University of Queensland}
  \city{Brisbane}
  \country{Australia}
}

\author{Ruiqi Zheng}
\email{ruiqi.zheng@uq.net.au}
\affiliation{%
  \institution{The University of Queensland}
  \city{Brisbane}
  \country{Australia}
}

\author{Lizhen Cui}
\email{clz@sdu.edu.cn}
\affiliation{%
  \institution{Shandong University}
  \city{Jinan}
  \country{China}
}

\author{Yuhui Shi}
\authornote{Corresponding Author}
\email{shiyh@sustech.edu.cn}
\affiliation{%
  \institution{Southern University of Science and Technology}
  \city{Shenzhen}
  \country{China}
}

\author{Hongzhi Yin}
\authornotemark[1]
\email{db.hongzhi@gmail.com}
\affiliation{%
  \institution{The University of Queensland}
  \city{Brisbane}
  \country{Australia}
}


\begin{abstract}
Federated recommender systems (FedRecs) have gained significant attention for their potential to protect user's privacy by keeping user privacy data locally and only communicating model parameters/gradients to the server. 
Nevertheless, the currently existing architecture of FedRecs assumes that all users have the same 0-privacy budget, i.e., they do not upload any data to the server, thus overlooking those users who are less concerned about privacy and are willing to upload data to get a better recommendation service. 
To bridge this gap, this paper explores a user-governed data contribution federated recommendation architecture where users are free to take control of whether they share data and the proportion of data they share to the server. 
To this end, this paper presents a cloud-device collaborative graph neural network federated recommendation model, named CDCGNNFed. It trains user-centric ego graphs locally, and high-order graphs based on user-shared data in the server in a collaborative manner via contrastive learning. 
Furthermore, a graph mending strategy is utilized to predict missing links in the graph on the server, thus leveraging the capabilities of graph neural networks over high-order graphs.
Extensive experiments were conducted on two public datasets, and the results demonstrate the effectiveness of the proposed method.

\end{abstract}



\maketitle

\section{Introduction}

\begin{figure}[t]
\centering
\includegraphics[width=0.4\textwidth]{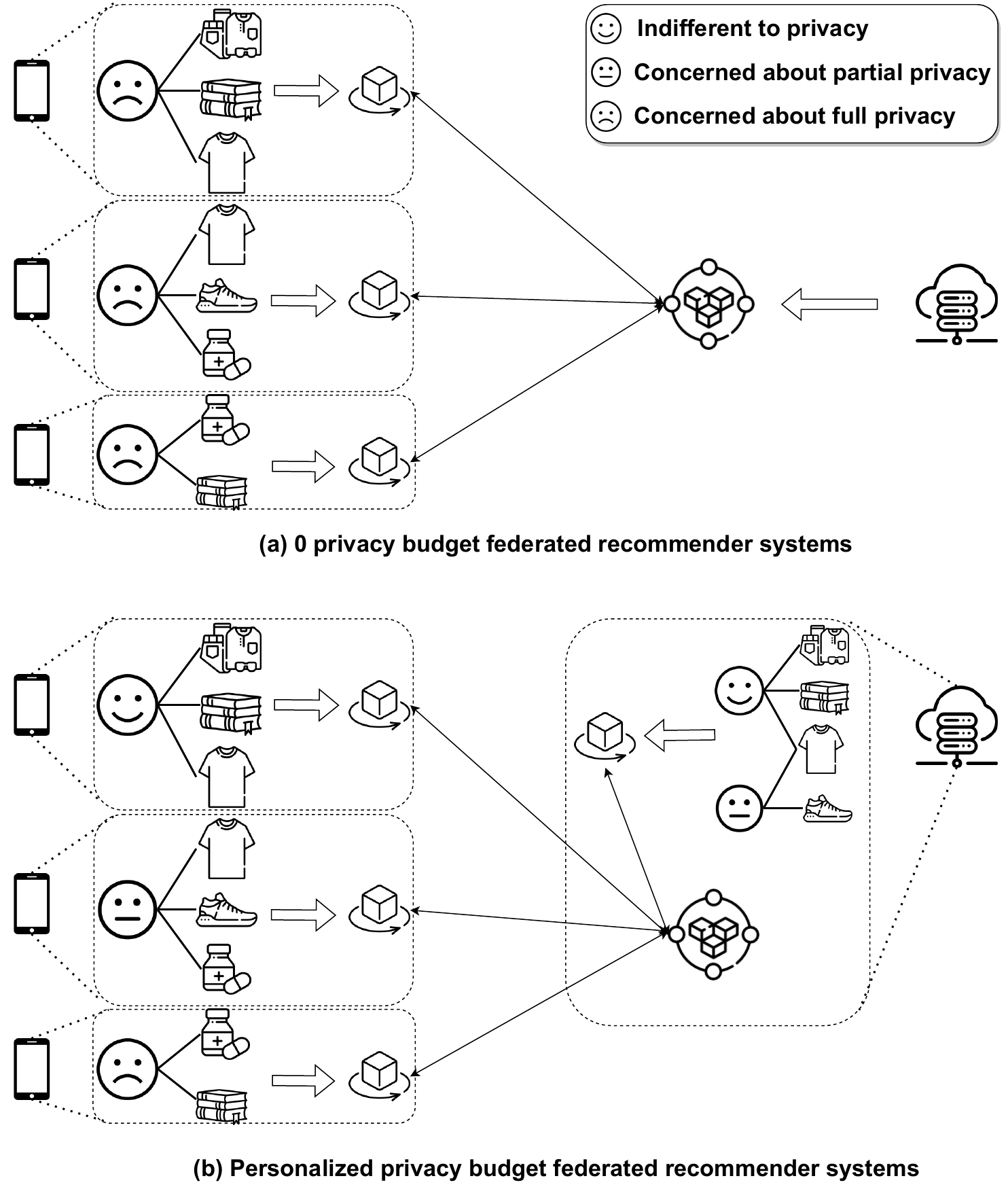} 
\caption{(a) 0 privacy budget federated recommender systems that users do not upload any data to the server. (b) Personalized privacy budget federated recommender systems that users are free to take control of whether they share
data and the proportion of data they share to the server.}
\label{fig:example}
\vspace{-15pt}
\end{figure}

Recommender systems \cite{bobadilla2013recommender,zhang2019deep,zheng2023automl,qu2022single} have been shown to be an effective technique for providing personalised content recommendation services (e.g., videos and goods) to users based on their preferences. Typically, the recommender system is deployed on a central server that collects all users' historical behavior data (e.g., clicks and purchases) to train a global recommendation model, and the more data that is collected, the more accurate the model is. However, such recommender systems inevitably raise privacy concerns due to their centralized data collection mechanism. Moreover, many regulations, such as the General Data Protection Regulation (GDPR\footnote{https://gdpr-info.eu/}), have recently been issued to better protect users' data privacy, so it is desirable to investigate how to balance privacy risks against recommendation utilities.

Recently, federated learning \cite{mcmahan2017communication,yang2019federated}, as a promising solution to privacy-preserving machine learning, has been widely adopted in recommender systems to mitigate privacy concerns, termed federated recommender systems \textit{ (FedRecs)} \cite{yang2020federated,sun2022survey}. Specifically, as shown in Figure \ref{fig:example} (a), the key idea of FedRecs is that all of the user's data is retained on their own device in a decentralized fashion. In each training round, the central server randomly selects a group of devices to train locally, and then only parameters/gradients (without actual data sharing) are aggregated to the central server to learn a global model that will be redistributed to each device. 
Research on FedRecs could be roughly classified into two categories: matrix factorization based FedRecs (MF-FedRecs) \cite{ammad2019federated,chai2020secure,lin2020fedrec} and graph neural networks based FedRecs (GNN-FedRecs) \cite{wu2022federated,10.1145/3501815,10.1145/3511808.3557668,qu2023semi}. MF-FedRecs mainly learns the global item embedding table by collaboratively training the local first-order user-item interaction matrix distributed across different devices. On the other hand, GNN-based recommender systems \cite{wu2022graph} have recently achieved state-of-the-art results due to their superior ability to effectively capture higher-order graph structural information compared to matrix factorization methods. However, in federated scenarios, each device has only a first-order user ego graph that includes the items that the user interacts with directly. Hence, the core challenge behind GNN-FedRecs is to learn higher-order graph structural information in a privacy-preserving manner. For example, FedGNN \cite{wu2022federated} presents to use a trusted third-party server to construct a high-order graph.

Nevertheless,  the currently existing architecture of FedRecs assumes that all users have the same 0-privacy budget, meaning they do not upload any data to the server, which is inflexible and unappealing due to the following reasons: (1) it overlooks those users who are less concerned about privacy and willing to share either all or portions of their data to receive a better recommendation service. (2) The model performance of FedRecs is generally degraded due to the non-identically distributed data among the users' devices \cite{karimireddy2019scaffold,li2019convergence}. Thus, adopting a uniform policy where all users are prohibited from uploading data could also hurt the revenue of platforms due to the degraded model performance. (3) It requires that users who need the recommendation service have to be involved in model training, which brings a huge burden to the user's device as it requires substantial computational and storage resources as well as communication costs.

To mitigate above issues, this paper explores a user-governed data contribution federated recommendation architecture, as shown in Figure \ref{fig:example} (b), where users are free to take control of whether they share data and the proportion of data they share to the server. In such a setting, this paper presents a cloud-device collaborative graph neural network federated recommendation model, named CDCGNNFed. It trains user-centric ego graphs locally, and high-order graphs based on user-shared data in the server in a collaborative manner via contrastive learning. Specifically, a graph mending strategy is first employed to predict missing links in the graph on the server, thus leveraging the capabilities of graph neural networks over high-order graphs. After that, for each training round, devices and the server independently infer and exchange embeddings, so that local and global views of the same node can be constructed as positive/negative pairs for contrastive learning.

Overall, our main contributions are summarized as follows:
\begin{itemize}
    \item To our best knowledge, this is the first work to investigate a more flexible and personalized privacy framework called user-governed data contribution federated recommendation (UGFedRec), where users have granular control over the extent to which they are willing to share data with the platform to balance privacy risks and recommendation utilities.
    \item In the UGFedRec setting, we propose a cloud-device collaborative graph neural network
    federated recommendation model, named CDCGNNFed, which trains user-centric ego graphs locally, and high-order graphs based on user-shared data in the server in a collaborative manner via contrastive learning.
    \item We conduct extensive experiments on public real-world datasets to validate the effectiveness of the proposed methods, and experimental results demonstrate that the proposed method can achieve a promising performance for the Top-K recommendation.
\end{itemize}

The remainder of this paper is organized as follows. Section 2 will review related work, and Section 3 will formulate the research problem and elaborate on the proposed method. The experiments are discussed in Section 4, followed by a conclusion in Section 5.

\section{Related Work}

\subsection{Centralized Recommendation}

Recommender systems \cite{yin2015joint} have been shown to be an effective technique for providing personalised content recommendation services (e.g., videos and goods) to users by collecting all users’ historical behavior data (e.g., clicks and purchases) to train a global recommendation model on the server. Methods in this field can be broadly categorised as MF-based methods, deep learning based methods and GNN-based methods. The main idea of MF-based methods \cite{koren2009matrix,ramlatchan2018survey} is to decompose the user-item interaction matrix into two lower-dimensional matrices representing latent features of users and items. Deep learning based methods \cite{covington2016deep,guo2017deepfm,yin2015joint} focus on leveraging deep neural networks to learn intricate patterns from user-item interaction data, often capturing non-linear relationships. In recent years, GNN-based recommender systems \cite{he2020lightgcn,ying2018graph,wu2022graph,zhang2021graph} have achieved state-of-the-art results due to their superior ability to effectively capture higher-order graph structural information. As previously highlighted, these methods are largely centralized, collecting user data for model training, which raises potential data privacy concerns.

\subsection{Federated Recommendation}
Drawing inspiration from the efficacy of federated learning in ensuring privacy in machine learning, FedRecs \cite{yin2024device,zhang2023comprehensive,yuan2023interaction,wang2022fast,yuan2023federated,yuan2023hetefedrec,yuan2023manipulating} have been introduced, allowing for cloud-device model collaborative training without actual data sharing. 
Research on FedRecs could be roughly classified into two categories: matrix factorization based FedRecs (MF-FedRecs) \cite{ammad2019federated,chai2020secure,lin2020fedrec} and graph neural networks based FedRecs (GNN-FedRecs) \cite{wu2022federated,10.1145/3501815,10.1145/3511808.3557668}. MF-FedRecs mainly learns the global item embedding table by collaboratively training the local first-order user-item interaction matrix distributed across different devices. For example, FCF \cite{ammad2019federated} extends collaborative centralized filtering to the federated model. In particular, it utilizes alternating least squares and stochastic gradient descent to optimize user and item embeddings on the device and server sides, respectively. On the other hand, GNN-based recommender systems \cite{wu2022graph} have recently achieved state-of-the-art results due to their superior ability to effectively capture higher-order graph structural information compared to matrix factorization methods. For instance, FedGNN \cite{wu2022federated} presents to use a trusted third-party server to construct high-order graph such that GNNs could be employed to learn user/item embeddings in a privacy-preserving manner.
Although currently FedRecs have attracted considerable interest in the privacy-preserving recommendation field, methods in the context of user-governed data contribution federated recommendation remain highly unexplored. A work similar to us is FedeRank \cite{anelli2021federank}, where users also have the ability to govern the proportion of data they upload. However, a distinguishing factor from our method is that in Federank, all user data remains local, while users can dictate the percentage of gradients corresponding to training samples that they transmit.

\section{Proposed Method}

\begin{figure*}[t]
\centering
\includegraphics[width=1\textwidth]{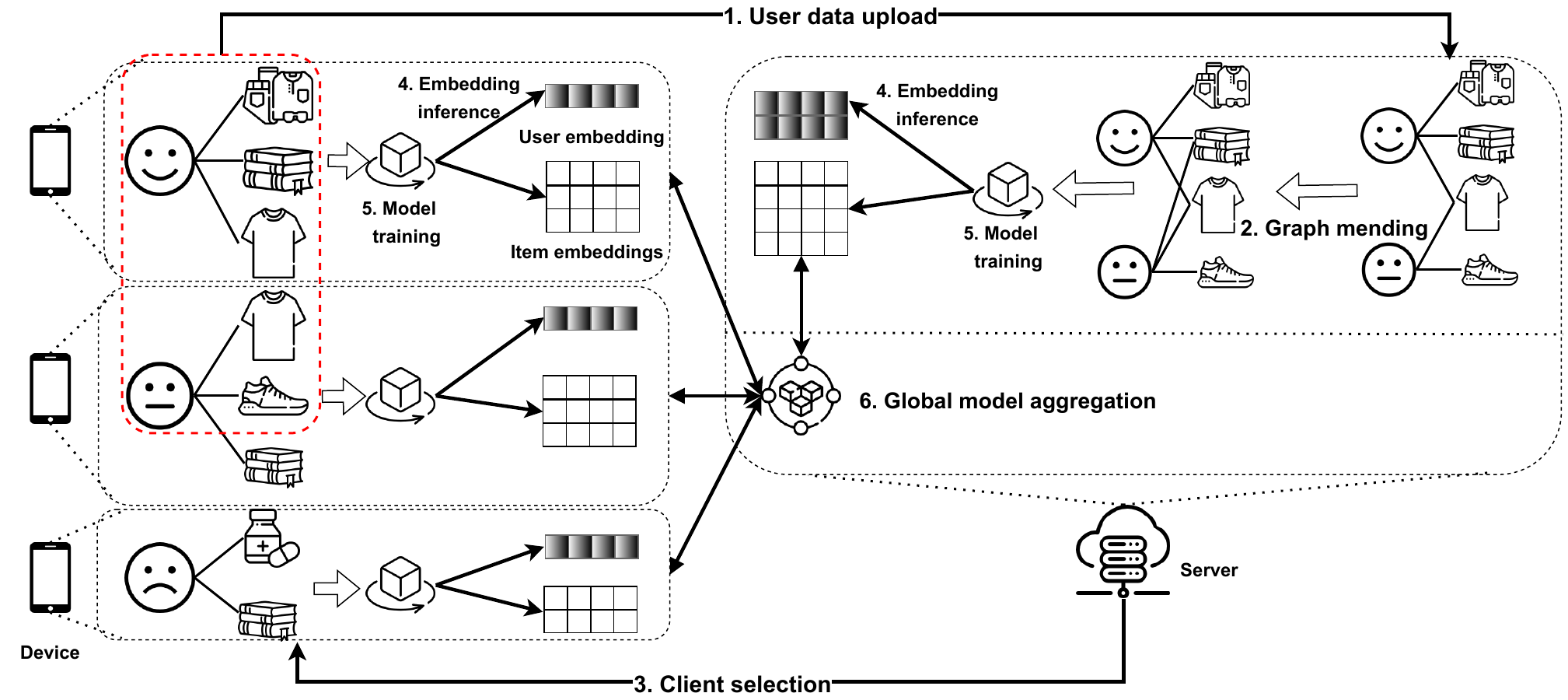} 
\caption{The architecture of the proposed method.}
\label{fig:overview}
\end{figure*}

In this section, we first formulate the research problem and then elaborate the proposed method.

\subsection{Problem formulation}
Let $\mathcal{U}$ and $\mathcal{I}$ represent a set of users/devices\footnote{We assume that each device is only associated with a single user. Therefore, we interchangeably use the terms ``device'' and ``user'' throughout this paper.} and items, respectively. $X \in \mathbb{R}^{|\mathcal{U}| \times |\mathcal{I}|}$ denotes the binary user-item interaction matrix where the element $x_{ui}$ represents the implicit feedback between user $u \in \mathcal{U}$ and item $i \in \mathcal{I}$. Specifically, $x_{ui} = 1$ and $x_{ui} = 0 $ indicate whether there is an interaction or not, respectively. In addition, the embedding-based recommendation model is denoted as $f(\Theta)$ parameterized by $\Theta$. Specifically, it maps users and items into a shared embedding space via the model $f(\Theta): \mathcal{U},\mathcal{I} \rightarrow P \in \mathbb{R}^{|\mathcal{U}| \times d}, Q \in \mathbb{R}^{|\mathcal{I}| \times d}$, where the user embedding $\textbf{p}_{u} \in P$ and the item embedding $\textbf{q}_{i} \in Q$ represent the $d$-dimensional vector representations of the user $u$ and the item $i$, respectively.

In the traditional federated recommendation setting, each user $u$ keeps all of their own interaction data $X_{u} \in \mathbb{R}^{|\mathcal{I}|}$ corresponding to the $u$-th row of $X$ on their local device for the purpose of privacy protection. In addition, each device $u$ maintains its local model consisting of a set of local parameters that have model parameters $\Theta_{u}$, the user embedding $\textbf{p}_{u} \in \mathbb{R}^{d}$, and the item embedding table $Q_{u} \in \mathbb{R}^{|\mathcal{I}| \times d}$. For each training round, the server selects a set of devices, denoted $\mathcal{U}_{s} \subseteq \mathcal{U}$, to train their models locally. Each local device typically uploads the locally trained parameters $\Theta_{u}$ and the item embedding table $Q_{u}$ or their corresponding gradients $\nabla\Theta_{u}$ and $\nabla Q_{u}$ to the server. After that, the server will train a global model using the collected parameters/gradients via the aggregation function, such as FedAvg \cite{mcmahan2017communication}, and then redistribute the global model to all devices.

Although the federated recommendation architecture mentioned above can protect users' privacy by keeping all users' data locally, it assumes that all users have the same 0-privacy budget, i.e., they do not upload any data to the server, thus overlooking those users (denoted as $\mathcal{U}^{+}$) who are less concerned about privacy and willing to share either all or portions of their data to receive a better recommendation service. To bridge this gap, this work aims to explore a more flexible federated recommendation framework, termed the User-Governed Data Contribution Federated recommending System (UGFedRec), where users are free to take control of whether they share data and the proportion of data they share with the server. Specifically, the main difference between UGFedRec and traditional FedRec is that each user has the option to upload all, some, or no data to the server.
In this way, the server can also train a model based on the data voluntarily uploaded by users.
Finally, the goal of UGFedRec is to minimize the following loss function $\mathcal{L}$:
\begin{equation}
    \mathcal{L} = \sum_{u \in \mathcal{U}} \mathcal{L}_{u} +\mathcal{L}_{s}
\end{equation}
where $\mathcal{L}_{u}$ and $\mathcal{L}_{s}$ are loss functions for local devices and the server, respectively.

\subsection{CDCGNNFed}
This work aims to explore a more flexible and personalized privacy framework for user-governed data contribution federated recommendation (UGFedRec). To this end, we introduce a cloud-device collaborative graph neural network federated recommendation model, named CDCGNNFed. The architecture of the proposed method is shown in Figure \ref{fig:overview}, which encompasses the following steps: (1) \textbf{User data upload}: Users voluntarily choose to upload all, some, or no data to the server. (2) \textbf{Graph mending:} The server employs a graph mending strategy to predict missing links. (3) \textbf{Client selection:} The server randomly selects a set of devices to participate in the current round of training. (4) \textbf{Embedding inference:} Both the chosen devices and the server use their local data and the data voluntarily uploaded by users, respectively, to perform embedding inference for the subsequent contrastive learning purposes. (5) \textbf{Device and server model training:} Devices and the server independently train their models based on their training data. (6) \textbf{Global model aggregation:} Finally, a standard federated learning model aggregation is performed.

\subsection{Graph mending}
Within the context of UGFedRec, users voluntarily contribute either all or a portion of their interaction data to the server. Consequently, the server constructs a user-item bipartite graph $\mathcal{G}=\{\mathcal{U}^{+}, \mathcal{I}, X^{+}\}$, where $\mathcal{U}^{+}$ is the set of users willing to share their data, $\mathcal{I}$ denotes the set of all items, and $X^{+} \in \mathbb{R}^{|\mathcal{U}^{+}| \times |\mathcal{I}|}$ denotes the user-item interaction data provided by users. 
Nevertheless, when the volume of user-contributed data is limited, the graph 
$\mathcal{G}$ on the server may be disjointed, consisting of several subgraphs representing isolated user-item interactions. As a result, simply applying a GNN directly to these subgraphs might not effectively capture higher-order graph structural information, potentially leading to suboptimal outcomes. To address this challenge, inspired by the approach in FedSage+ \cite{zhang2021subgraph}, we introduce a graph mending strategy that predicts the missing links in graph $\mathcal{G}$. This enables the GNN to fully exploit its node representation capability, ensuring more comprehensive information capture from the graph structure. Specifically, 
we first employ a graph impairing strategy to extract a subset of links $\hat{X^{+}} \subset X^{+}$ serving as the ground truth for training graph mending by simulating the scenario wherein links may be missing. With the impaired links in place, we can leverage standard GNNs for learning the node embedding over the impaired graph $\hat{\mathcal{G}}$ as follows:
\begin{equation}
    Z_{u}, Z_{i} = GNN(\hat{\mathcal{G}})
\end{equation}
where $\textbf{z}_{u} \in Z_{u}, \textbf{z}_{i} \in Z_{i}$ are the user embedding and the item embedding, respectively. $GNN(\cdot)$ denotes GNN-based node encoder models. Since $\hat{\mathcal{G}}$ is a user-item bipartite graph in this context, we adopt Light Graph Convolution (LGC \cite{he2020lightgcn}) as the encoder, which will be provided more detailed introduction in the subsequent sections. In this way, for the pair of user-item nodes involved in the impaired link $e_{ui} \in \hat{X^{+}}$, we can use cosine similarity (denoted $cos(\cdot)$) to calculate the distance between the pair of nodes, and update the user/item embeddings using the following loss function $\mathcal{L}_{gm}$:
\begin{equation}
    \mathcal{L}_{gm} = \sum_{e_{ui} \in \hat{X^{+}}} ||cos(\textbf{z}_{u},\textbf{z}_{i})-e_{ui}||_{2}, e_{ui}=1, e_{ui} \in \hat{X^{+}}; e_{ui}=0, e_{ui} \notin \hat{X^{+}}
    \label{equ:missing}
\end{equation}
Finally, we can predict the potential missing links by calculating the cosine similarity of user-item node pairs of $\mathcal{G}$, and determine whether to establish a link by comparing it to a predefined threshold $t$.

\subsection{Embedding inference}
After predicting the missing links on the server side, we perform embedding inference separately on the device side and server side, aiming to obtain the local view of the same node on the device side and the global view on the server side.

For the device side embedding inference, since each device has only a first-order ego graph that includes the items that the user $u$ interacts with directly. Thus, the user/item embeddings (i.e., $\mathbf{e}_{u-d}^{(l+1)}$,$\mathbf{e}_{i-d}^{(l+1)}$) at the $(l+1)$-th layer could be learned by LGC \cite{he2020lightgcn} as follows:
\begin{equation}
\begin{aligned}
&\mathbf{e}_{u-d}^{(l+1)}=\sum_{i \in \mathcal{N}(u)} \frac{1}{\sqrt{\left|\mathcal{N}(u)\right|} \sqrt{\left|\mathcal{N}(i)\right|}} \mathbf{e}_{i-d}^{(l)} \\
&\mathbf{e}_{i-d}^{(l+1)}=\sum_{u \in \mathcal{N}(i)} \frac{1}{\sqrt{\left|\mathcal{N}(i)\right|} \sqrt{\left|\mathcal{N}(u)\right|}} \mathbf{e}_{u-d}^{(l)}
\end{aligned}
\label{equ:LGC}
\end{equation}
Notably, on each device, a user has only a first-order ego graph, which includes the user and the items they have interacted with. This means for each item $i$, its neighborhood $\mathcal{N}(i)=\{u\}$ consists only of the user $u$, and hence $|\mathcal{N}(i)|=1$. Finally, we can use layer combination method \cite{he2020lightgcn} to obtain the final user/item embeddings, denoted as $\textbf{e}_{u-d}$ and $\textbf{e}_{i-d}$, on the device side as below: 
\begin{equation}
\mathbf{e}_{u-d}=\sum_{l=0}^L \alpha_l \mathbf{e}_{u-d}^{(l)} ; \quad \mathbf{e}_{i-d}=\sum_{l=0}^L \alpha_l \mathbf{e}_{i-d}^{(l)}
\label{equ:layercomb}
\end{equation}
where  $\alpha_l$ is the hyperparameter representing the importance of the $l$-th layer. Since the graph on the device side is first-order, we are limited to using a single layer of LGC, i.e., $L=1$.

For the server side embedding inference, the graph on the server side possesses higher-order graph structural information after graph mending. Similarly, we can calculate the embeddings of the user and items at each layer by Equation \ref{equ:LGC}, and then obtain the final user/item embeddings $\textbf{e}_{u-s}$ and $\textbf{e}_{i-s}$ by Equation \ref{equ:layercomb} on the server side.

\subsection{Device-server constrastive learning}
After obtaining the user/item embeddings on both the local device and the server, users who willingly share ALL their data are assumed to be less concerned about their privacy. These users compute their user embeddings locally and upload them to the server. The server, in turn, distributes these embeddings to devices involved in each training round for the subsequent contrastive learning purposes. Users who share only PART of their data are assumed to be somewhat privacy-conscious. They upload their locally computed user embeddings to the server. However, the server does not distribute their embeddings to other devices. Users who choose not to share their data are considered highly privacy-conscious. Their user embeddings are strictly retained locally, not uploaded to the server, nor shared with other devices.
As a result, for the same user/item node, we can consider the embedding obtained from the local device as the node's local view, and the embedding from the server as its global view. In this way, we can construct a positive pair, that is, $\{(\textbf{e}_{u-d},\textbf{e}_{u-s}) | u \in \mathcal{U}^{+}\}$ for contrastive learning, and treat views from different nodes as negative pairs, that is, $\{(\textbf{e}_{u-d},\textbf{e}_{v-s}) | u,v \in \mathcal{U}^{+}, u \neq v \}$. Formally, we follow SimCLR \cite{chen2020simple} to adopt contrastive loss InfoNCE \cite{gutmann2010noise} as follows:
\begin{equation}
    \mathcal{L}_{CL}^{user} = \sum_{u \in \mathcal{U}^{+}}-log\frac{exp(cos(\textbf{e}_{u-d},\textbf{e}_{u-s})/\tau)}{\sum_{v \in \mathcal{U}^{+}exp(cos(\textbf{e}_{u-d},\textbf{e}_{v-s})/\tau)}}
\end{equation}
where $\tau$ is the hyperparameter known as temperature. Analogously, the contrastive loss of items is denoted as $\mathcal{L}_{CL}^{item}$. In this way, the final contrastive learning loss is $\mathcal{L}_{CL} = \mathcal{L}_{CL}^{user}+\mathcal{L}_{CL}^{item}$ \cite{wu2021self}. 

\subsection{Device and server model training}
In the context of local model updates for devices, for users who are not willing to upload their data to the server, we leverage the data available on their local devices and update using the Bayesian Personalized Ranking (BPR) \cite{rendle2012bpr} loss function $\mathcal{L}_{BPR}$ as follows:.
\begin{equation}
\mathcal{L}_{BPR}=- \sum_{i \in \mathcal{N}_u} \sum_{j \notin \mathcal{N}_u} \ln \sigma\left(cos(\textbf{e}_{u-d},\textbf{e}_{i-d})-cos(\textbf{e}_{u-d},\textbf{e}_{j-d})\right)+\lambda\left\|\mathbf{e}\right\|^2
\label{equ:bpr}
\end{equation}
where 
$\lambda$ is a hyperparameter for controlling the strength of the $L_{2}$ regularization. Analogously, for those users who willing to share their data with the server and for the server-side model, we train models using a combination of the BPR loss and contrastive learning loss, as illustrated below.
\begin{equation}
    \mathcal{L} =\mathcal{L}_{BPR} +\lambda_{1} \mathcal{L}_{CL} + \lambda\left\|\mathbf{e}\right\|^2
\label{equ:bpr+ssl}
\end{equation}
where $\lambda_{1}$ and $\lambda$ are hyperparameters for controlling the strength of contrastive learning loss and regularization. 
In an extreme situation where no users are willing to share all their data, there may still be users who are willing to share part of their data and upload their user embeddings to the server. As per our user embedding management strategy mentioned earlier, the server, in this case, will not distribute any user embeddings to the devices. Consequently, only the server will be able to calculate the contrastive learning loss, while local training will be limited to using the BPR loss.

\subsection{Global model aggregation}
Finally, after both the devices and the server have completed their model updates, we can implement a standard global model aggregation process, similar to that used in federated learning. In this process, the model on the server is treated as a special device. Techniques such as Local Differential Privacy (LDP \cite{dwork2006calibrating}) can be used to encrypt the model parameters uploaded by the devices, ensuring data privacy and security. Subsequently, the FedAvg \cite{mcmahan2017communication} algorithm can be employed to aggregate these parameters to form a global model.


    



\section{Experiments}
In this section, we will first introduce the experimental settings, and then report and discuss the experimental results for answering the following research questions:
\begin{itemize}
    \item How does the proposed method compare with other federated recommendation methods in the UGFedRec setting?
    \item How do various components, such as contrastive learning, influence the performance of the proposed method?  
    \item How do various hyperparameters influence the performance of the proposed method?
\end{itemize}

\subsection{Settings}

\subsubsection{Datasets}

To validate the effectiveness of the proposed method, the experiment is carried out on two public datasets, including Gowalla \cite{liang2016modeling}, and Yelp2018\footnote{https://www.yelp.com/dataset/challenge}. The statistics of datasets are described in Table \ref{tab:datasets}. The Gowalla is a location-based social network dataset consisting of users and their locations by checking-in. On the other hand, the Yelp2018 dataset is a business review dataset that includes customers, restaurants, and the associated reviews given by customers to these restaurants. Following \cite{10.1145/3038912.3052569,KGAT19}, we exclude users and items from Gowalla and Yelp2018 that have fewer than 20 and 10 interactions, respectively. Each of the three datasets is then partitioned into training, validation, and test sets in an 8:1:1 ratio, respectively.

\begin{table}[htbp]
  \centering
  \caption{The statistics of datasets.}
    \begin{tabular}{l|c|c|c}
    \toprule
    Datasets & \multicolumn{1}{l|}{\#Users} & \multicolumn{1}{l|}{\#Items} & \multicolumn{1}{l}{\#Interactions} \\
    \midrule
    \midrule
    Gowalla & 29858  & 40981  & 1027370  \\
    Yelp2018  & 31668  & 38048  & 1561406  \\
    \bottomrule
    \end{tabular}%
  \label{tab:datasets}%
\end{table}%

\subsubsection{Baselines:}
\begin{table*}[htbp]
  \centering
  \caption{The model performance with respect to Recall@20 and NDCG@20 on Gowalla for the partial uploading case.}
    \begin{tabular}{clccccccccccc}
    \toprule
    \multicolumn{2}{c}{Share ratio} & (0,0.1) & [0.1,0.2) & [0.2,0.3) & [0.3,0.4) & [0.4,0.5) & [0.5,0.6) & [0.6,0.7) & [0.7,0.8) & [0.8,0.9) & [0.9,1) & [0,1]\\
    \midrule
    \midrule
    \multirow{4}[2]{*}{Recall@20} & FedeRank & 0.1438 & 0.1443 & 0.145 & 0.1453 & 0.1461 & 0.1468 & 0.1474 & 0.1479 & 0.1484 & 0.1491 & 0.1477 \\
          & UGFed-MF & 0.1442 & 0.145 & 0.145 & 0.1463 & 0.1471 & 0.1478 & 0.1486 & 0.1493 & 0.15  & 0.1502 & 0.1489 \\
          & UGFed-GNN & 0.1453 & 0.146 & 0.1468 & 0.1498 & 0.1553 & 0.1573 & 0.1618 & 0.1704 & 0.1705 & 0.1752 & 0.1716\\
          & \textbf{CDCGNNFed} & \textbf{0.1463} & \textbf{0.1478} & \textbf{0.1502} & \textbf{0.1531} & \textbf{0.1556} & \textbf{0.1583} & \textbf{0.1704} & \textbf{0.1727} & \textbf{0.1778} & \textbf{0.1809} & \textbf{0.1724} \\
    \midrule
    \multirow{4}[2]{*}{NDCG@20} & FedeRank & 0.1206 & 0.1215 & 0.1228 & 0.1236 & 0.1247 & 0.1253 & 0.1268 & 0.1274 & 0.1287 & 0.1294 & 0.1269 \\
          & UGFed-MF & 0.1227 & 0.1235 & 0.1241 & 0.125 & 0.1258 & 0.1263 & 0.1272 & 0.128 & 0.1295 & 0.1302 & 0.127\\
          & UGFed-GNN & 0.1163 & 0.119 & 0.1216 & 0.1245 & 0.1272 & 0.1301 & 0.1328 & 0.1356 & 0.1383 & 0.141 & 0.1396\\
          & \textbf{CDCGNNFed} & \textbf{0.1228} & \textbf{0.1236} & \textbf{0.1244} & \textbf{0.1278} & \textbf{0.131} & \textbf{0.1345} & \textbf{0.1379} & \textbf{0.1452} & \textbf{0.1481} & \textbf{0.154} & \textbf{0.1448}\\
    \bottomrule
    \end{tabular}%
  \label{tab:addlabel}%
\end{table*}%

\begin{table*}[htbp]
  \centering
  \caption{The model performance with respect to Recall@20 and NDCG@20 on Yelp 2018 for the partial uploading case.}
    \begin{tabular}{clccccccccccc}
    \toprule
    \multicolumn{2}{c}{Share ratio} & (0,0.1) & [0.1,0.2) & [0.2,0.3) & [0.3,0.4) & [0.4,0.5) & [0.5,0.6) & [0.6,0.7) & [0.7,0.8) & [0.8,0.9) & [0.9,1) & [0,1] \\
    \midrule
    \midrule
    \multirow{4}[2]{*}{Recall@20} & FedeRank & 0.0482 & 0.0491 & 0.0498 & 0.0506 & 0.0511 & 0.0521 & 0.0527 & 0.0533 & 0.0541 & 0.0562 & 0.0499 \\
          & UGFed-MF & 0.0486 & 0.0493 & 0.0498 & 0.0504 & 0.0512 & 0.0521 & 0.0529 & 0.0542 & 0.0556 & 0.0571 & 0.0527\\
          & UGFed-GNN & 0.0461 & 0.0475 & 0.0489 & 0.051 & 0.0523 & 0.0541 & 0.0558 & 0.0574 & 0.0597 & 0.0615 & 0.056\\
          & \textbf{CDCGNNFed} & \textbf{0.0493} & \textbf{0.0505} & \textbf{0.0518} & \textbf{0.0531} & \textbf{0.0544} & \textbf{0.0557} & \textbf{0.0568} & \textbf{0.0582} & \textbf{0.0605} & \textbf{0.0623} & \textbf{0.0579} \\
    \midrule
    \multirow{4}[2]{*}{NDCG@20} & FedeRank & 0.041 & 0.0414 & 0.0419 & 0.0423 & 0.0428 & 0.0433 & 0.0438 & 0.0444 & 0.0451 & 0.046 & 0.0433 \\
          & UGFed-MF & 0.0415 & 0.042 & 0.0427 & 0.0432 & 0.0429 & 0.0435 & 0.0442 & 0.0448 & 0.0458 & 0.0467 & 0.0431\\
          & UGFed-GNN & 0.0423 & 0.0432 & 0.0442 & 0.0454 & 0.0462 & 0.0475 & 0.0484 & 0.0495 & 0.0497 & 0.045 & 0.0478\\
          & \textbf{CDCGNNFed} & \textbf{0.0432} & \textbf{0.0444} & \textbf{0.045} & \textbf{0.0461} & \textbf{0.0468} & \textbf{0.0479} & \textbf{0.0485} & \textbf{0.0498} & \textbf{0.0507} & \textbf{0.0515} & \textbf{0.0491}\\
    \bottomrule
    \end{tabular}%
  \label{tab:addlabel}%
\end{table*}%

\begin{table*}[htbp]
  \centering
  \caption{The model performance with respect to Recall@20 and NDCG@20 on Gowalla and Yelp 2018 for the no uploading and full uploading cases.}
    \begin{tabular}{cl|cccc|cccc}
    \toprule
    \multicolumn{2}{c|}{\multirow{2}[4]{*}{Datasets}} & \multicolumn{4}{c|}{Gowalla}  & \multicolumn{4}{c}{Yelp2018} \\
\cmidrule{3-10}    \multicolumn{2}{c|}{} & \multicolumn{2}{c}{Recall@20} & \multicolumn{2}{c|}{NDCG@20} & \multicolumn{2}{c}{Recall@20} & \multicolumn{2}{c}{NDCG@20} \\
    \midrule
    \multicolumn{2}{c|}{Share ratio} & 0     & 1     & 0     & 1     & 0     & 1     & 0     & 1 \\
    \midrule
    \midrule
    \multirow{2}[2]{*}{Cloud} & NeuMF & -     & 0.1509 & -     & 0.1309 & -     & 0.0586 & -     & 0.0476 \\
          & LightGCN & -     & 0.1811 & -     & 0.1534 & -     & 0.0627 & -     & 0.0509 \\
    \midrule
    \multirow{2}[2]{*}{FedRec} & FedMF & 0.1435 & -     & 0.122 & -     & \textbf{0.0482} & -     & \textbf{0.0413} & - \\
          & FedPerGNN & \textbf{0.1442} & -     & \textbf{0.1233} & -     & 0.0453 & -     & 0.0409 & - \\
    \midrule
    \multirow{4}[2]{*}{UGFedRec} & FedeRank & 0.1432 & 0.1494 & 0.1197 & 0.1308 & 0.0476 & 0.0574 & 0.0408 & 0.047 \\
          & UGFed-MF & 0.1435 & 0.1504 & 0.122 & 0.1312 & \textbf{0.0482} & 0.0583 & \textbf{0.0413} & 0.0475 \\
          & UGFed-GNN & 0.1428 & 0.1776 & 0.1117 & 0.1538 & 0.0457 & 0.0626 & 0.0412 & 0.0503 \\
          & \textbf{CDCGNNFed} & 0.1428 & \textbf{0.1823} & 0.1117 & \textbf{0.1553} & 0.0457 & \textbf{0.0639} & 0.0412 & \textbf{0.0522} \\
    \bottomrule
    \end{tabular}%
  \label{tab:addlabel}%
\end{table*}%

\begin{itemize}
    \item \textbf{Cloud-based recommendation methods:}
    \begin{itemize}
        \item \textbf{NeuMF} \cite{10.1145/3038912.3052569}: It is the state-of-the-art MF-based deep recommendation method, utilizing DNN to supplant the dot product function, thereby capturing the non-linearity present in implicit feedbacks.
        \item \textbf{LightGCN} \cite{he2020lightgcn}: It is the state-of-the-art GNN-based recommendation method, utilizing GNN to capture high-order graph structure information via the linear neighborhood aggregation mechanism.
    \end{itemize}
    \item \textbf{FedRecs:}
    \begin{itemize}
        \item \textbf{FedMF} \cite{chai2020secure}: It is a MF-FedRec method which introduces a user-centric distributed matrix factorization framework, leveraging the homomorphic encryption technique to ensure users' privacy.
        \item \textbf{FedPerGNN} \cite{wu2022federated}: It is GNN-FedRec method which employs 
        a trusted third-party server to allocate neighbors, who share co-interacted items, to individual users, thereby leveraing the capabilities of GNN on capturing the high-order graph information.
    \end{itemize}
    \item \textbf{UGFedRecs}
    \begin{itemize}
    \item \textbf{FedeRank \cite{anelli2021federank}}: It is a MF-based UGFedRec method. In contrast to our approach, this method still retains all user data locally. However, it allows users to control the proportion of gradients corresponding to the training samples that are uploaded to the server.
    \item \textbf{UGFed-MF, and UGFed-GNN:} In the UGFedRec setting, the most naive approach would be to treat the server, collecting data voluntarily uploaded by users, as another device equivalent to other user devices, and then proceed with standard federated learning. Thus, we adopt MF and GNN as base models respectively, denoted as UGFed-MF and UGFed-GNN, to serve as baselines under this setting. We consider items that a user hasn't interacted with as potential candidates and report the results averaged across all users.
\end{itemize}
\end{itemize}

\subsubsection{Evaluation Metrics}
To evaluate the model performance, we employ two commonly used metrics, i.e., Recall@20 and NDCG@20 (Normalized Discounted Cumulative Gain) throughout experiments \cite{he2020lightgcn,10.1145/3038912.3052569}. The former measures the proportion of relevant items found within the top-20 recommendations, and the latter evaluates not only the presence of relevant items in the top-20 but also their ranking quality, with higher positions being more valuable. Following \cite{he2020lightgcn}, we consider items that a user hasn't interacted with as potential candidates and report the results averaged across all users.

\subsubsection{Hyper-parameter Settings}
We employ Xavier method \cite{glorot2010understanding} to initialize user and item embeddings with the embedding dimension 64 for all methods. We use Adam \cite{kingma2014adam} as the optimizer, and the learning rate and weight decay are search from $\{0.001,0.0005,0.0001\}$ and $\{0.005, 0.0001, 0.0005, 0.00001\}$ via grid search, respectively. In addition, the number of devices sampled for each training round is 256 and 528 for Gowalla and Yelp2018, respectively. The number of GNN layers for devices and the server models is 1 and 3, respectively. We will discuss settings of other hyperparameters in section 4.4. The baselines are implemented by the codes provided by the authors.

\subsection{Top-K Recommendation (RQ1)}
We first validate the effectiveness of our method on the prevalent top-k recommendation task commonly seen in recommendation systems. Our evaluation initially simulates scenarios where users have autonomy over data uploading, encompassing three distinct cases: (1) No uploading (i.e., sharing ratio of 0), which renders the model equivalent to a traditional federated recommendation system with a 0-privacy budget; (2) Partial uploading (with a sharing ratio between 0 and 1). Here, we randomly sample a value between [0,1] for each user to represent the proportion of their data shared. It's important to note that since it's numerically improbable to exactly hit the boundaries of 0 and 1, we approximate users with probabilities in the range [0,0.05] as No uploading users, meaning they share nothing. Similarly, users in the range [0.95,1] are considered as Full uploading, i.e., sharing all their data. (3) Full uploading (sharing ratio of 1), under which circumstance the model aligns with a centralized recommendation system. For each case, we independently execute the model five times using different random seeds and report the averaged outcomes, and all results are statistically significant with $p < 0.05$. The results for the partial data uploading scenario are delineated in Tables 2 and 3 for two distinct datasets, while results for no uploading and full uploading scenarios are reported in Table 4. From the results, we can observer that:

\begin{itemize}
    \item Overall, our proposed method outperforms baselines in the majority of cases, attesting to the effectiveness of the approach we've introduced.
    \item In most cases, GNN-based methods outperform those built on MF. Notably, for partial uploaded scenarios with smaller sharing ratios, the advantage of GNN is less pronounced. This can possibly be attributed to the server collating predominantly independent lower-order graphs, thereby mitigating GNN's potential in capturing higher-order graph structures. However, as the sharing ratio increases, the GNN-based techniques significantly overshadow MF-based methods.
    \item For the partial uploading case, our method, in tandem with UGFed-MF and UGFed-GNN, frequently outperforms Federank. This observation is plausible since, unlike Federank, we directly upload data to the server, rather than transmitting select gradients.
    \item In the no uploading scenarios, our model's performance aligns closely with the standard federated model. The MF-based approaches yield relatively better results. This behavior is understandable as there is no supplementary data available for utilization, leading our system to revert to the conventional federated recommendation model. Conversely, the GNN-based method, which can only harness first-order graph information, results in a somewhat suboptimal outcome.
    \item For full uploading scenarios, our model exhibits superior performance compared to centralized approaches. We attribute this enhanced performance partly to the introduced contrastive learning component. Moreover, by predicting missing links, our method partly alleviates the data sparsity issue, which subsequently enhances model performance.
\end{itemize}

\subsection{Ablation Study (RQ2)}
In this section, we aim to demonstrate the effect of the graph mending strategy component for predicting missing links and the device-cloud constrastive learning component for learning local and global views for the same node. To this end, we implement CDCGNNFed without graph mending strategy and constrastive learning, denoted as \textit{ w/o GM}
and \textit{ w/o CL}, respectively. The experiments are carried out on two datasets, and other settings are the same as the partial uploading scenario with share ratio between [0,1]. Experimental results are reported in Table \ref{tab:ablation}, from which we can observe that:

\begin{itemize}
    \item Overall, the removal of any component results in a significant deterioration in model performance. This underscores the indispensability and efficacy of both components within the model.
    \item Notably, the removal of the GM component leads to a more pronounced degradation in performance for both datasets. A possible explanation for this is the sparsity of the two datasets. The GM component proves adept at predicting missing links, effectively mitigating the challenges posed by this sparsity.
    \item The performance also sees a marked decline upon the removal of the CL component. This reiterates the potency of leveraging the contrastive learning component in rendering the learned embeddings more expressive. Concurrently, it affirms that in the UGFedRec context, the naive approach of viewing the server as a specialized device is suboptimal.
\end{itemize}

\begin{table}[htbp]
  \centering
  \caption{Ablation studies results with respect to GG and CL components.}
    \begin{tabular}{l|rr|rr}
    \toprule
    \multicolumn{1}{c|}{\multirow{2}[4]{*}{Method}} & \multicolumn{2}{c|}{Gowalla} & \multicolumn{2}{c}{Yelp2018} \\
\cmidrule{2-5}          & \multicolumn{1}{l}{Recall@20} & \multicolumn{1}{l|}{NDCG@20} & \multicolumn{1}{l}{Recall@20} & \multicolumn{1}{l}{NDCG@20} \\
    \midrule
    \midrule
    w/o GM & 0.1697 & 0.1441 & 0.0525 & 0.0413 \\
    w/o CL & 0.171 & 0.1443 & 0.053 & 0.0434 \\
    CDCGNNFed & \textbf{0.1724} & \textbf{0.1448} & \textbf{0.0579} & \textbf{0.0491} \\
    \bottomrule
    \end{tabular}%
  \label{tab:ablation}%
\end{table}%

\subsection{Hyperparameter analysis (RQ3)}
\begin{figure}[t]
\centering
\includegraphics[width=0.5\textwidth]{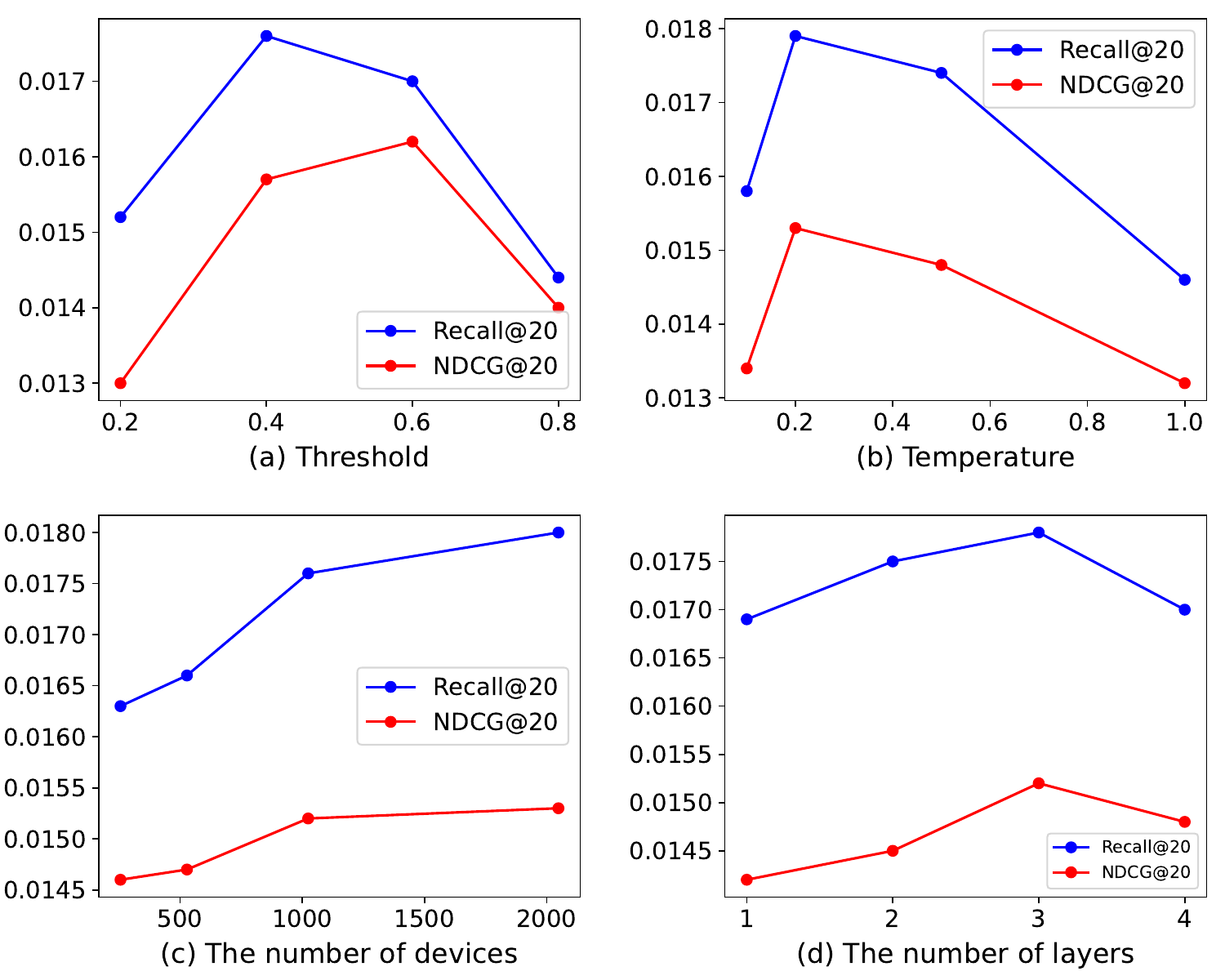} 
\caption{The performance of model with different hyper-parameter settings on Gowalla dataset. (a) Threshold $t$; (b) Temperature $\tau$; (c) The number of devices $|\mathcal{U}_{s}|$; (d) The number of layers $l$.}
\label{fig:hyper}
\vspace{-15pt}
\end{figure}

In this section, we investigate the impact of the four critical hyperparameters associated with our proposed method on the model's performance on Gowalla dataset. These hyperparameters include: (1) Threshold $t=\{0.2,0.4,0.6,0.8\}$ for graph mending strategy. (2) Temperature $\tau=\{0.1,0.2,0.5,1\}$ for constrastive learning; (3) The number of devices $|\mathcal{U}_{s}|=\{256,528,1024,2048\}$ for each training round; (4) The number of layers $l=\{1,2,3,4\}$ for GNN. The experimental settings are the same as the partial uploading scenario, and results are shown in Figure \ref{fig:hyper}. We can observe that: 

\begin{itemize}
    \item As the threshold increases, the model's performance initially rises but subsequently declines. This observation is rational, stemming from the fact that at lower thresholds, the graph mending strategy tends to produce a higher number of links, potentially introducing false negative links that degrade model performance. On the contrary, at higher thresholds, fewer links are generated, resulting in the persistence of numerous isolated subgraphs at the server end. This situation impedes the full exploitation of the Graph Neural Network's (GNN) inherent capabilities, leading to a drop in model performance.
    \item With regard to the temperature parameter in contrastive learning, the model's performance significantly deteriorates when the parameter's value is small. A plausible explanation for this decline is that the model's optimization process is dominated by the negatives. On the other hand, when the temperature parameter has a larger value, the model also does not perform optimally. This could be attributed to the model requiring a greater number of epochs to converge.
    \item As the number of devices participating in training increases per round, the model's performance gradually improves and stabilizes. A potential explanation for this improvement is that with a growing number of participating devices, there is a greater probability of incorporating users who actively share their data in each round. This, in turn, enhances the proportion of the model utilizing the contrastive learning strategy, thereby boosting its overall performance.
    \item The model exhibits optimal performance when the depth of the GNN is set to three layers. A plausible reason for this is that with fewer layers, the GNN may not capture the higher-order graph structural information effectively. On the other hand, when the network is too deep, it might encounter the over-smoothing issue \cite{li2018deeper}, consequently diminishing the model's efficacy.
\end{itemize}

\section{Conclusion}
In this study, we contend that the prevailing FedRecs architecture lacks adaptability and is less enticing. This is primarily because it uniformly assumes a 0-privacy budget for all users. Such an assumption fails to account for those individuals who, being less privacy-conscious, are open to sharing either their complete data or parts of it in exchange for enhanced recommendation services. To address this concern, we delve into a largely untapped area termed as the user-governed data contribution federated recommendation (UGFedRec). This paradigm empowers users with the autonomy to decide if they want to share data and, if so, the extent to which they would share with the server. Building on this concept, we introduce a cloud-device collaborative graph neural network federated recommendation model, dubbed CDCGNNFed. This model facilitates the training of user-centric ego graphs at the local level, while also leveraging high-order graphs constructed from user-contributed data on the server. The collaboration between the two is further enriched through contrastive learning. The efficacy of our proposed approach was validated on two public datasets. The experimental results demonstrate that, within the context of UGFedRec settings, our model consistently outperforms the existing baselines in the vast majority of scenarios. 
In future work, we intend to explore the integration of our framework with various base recommendation models, such as MF-based recommendation techniques. Additionally, we aim to address the cold-start problem inherent in federated recommender systems.

\section{ACKNOWLEDGEMENT}
This work is supported by the Australian Research Council under the streams of Future Fellowship (Grant No. FT210100624), the Discovery Project (Grants No. DP240101108), the Shenzhen Fundamental Research Program under Grant No. JCYJ20200109141235597, the National Science Foundation
of China under Grant No. 61761136008, the Shenzhen Peacock Plan under Grant No. KQTD2016112514355531, and the Program for Guangdong Introducing Innovative
and Entrepreneurial Teams under Grant No. 2017ZT07X386.

\bibliographystyle{ACM-Reference-Format}
\bibliography{main}


\begin{thebibliography}{46}


\ifx \showCODEN    \undefined \def \showCODEN     #1{\unskip}     \fi
\ifx \showDOI      \undefined \def \showDOI       #1{#1}\fi
\ifx \showISBNx    \undefined \def \showISBNx     #1{\unskip}     \fi
\ifx \showISBNxiii \undefined \def \showISBNxiii  #1{\unskip}     \fi
\ifx \showISSN     \undefined \def \showISSN      #1{\unskip}     \fi
\ifx \showLCCN     \undefined \def \showLCCN      #1{\unskip}     \fi
\ifx \shownote     \undefined \def \shownote      #1{#1}          \fi
\ifx \showarticletitle \undefined \def \showarticletitle #1{#1}   \fi
\ifx \showURL      \undefined \def \showURL       {\relax}        \fi
\providecommand\bibfield[2]{#2}
\providecommand\bibinfo[2]{#2}
\providecommand\natexlab[1]{#1}
\providecommand\showeprint[2][]{arXiv:#2}

\bibitem[Ammad-Ud-Din et~al\mbox{.}(2019)]%
        {ammad2019federated}
\bibfield{author}{\bibinfo{person}{Muhammad Ammad-Ud-Din}, \bibinfo{person}{Elena Ivannikova}, \bibinfo{person}{Suleiman~A Khan}, \bibinfo{person}{Were Oyomno}, \bibinfo{person}{Qiang Fu}, \bibinfo{person}{Kuan~Eeik Tan}, {and} \bibinfo{person}{Adrian Flanagan}.} \bibinfo{year}{2019}\natexlab{}.
\newblock \showarticletitle{Federated collaborative filtering for privacy-preserving personalized recommendation system}.
\newblock \bibinfo{journal}{\emph{arXiv preprint arXiv:1901.09888}} (\bibinfo{year}{2019}).
\newblock


\bibitem[Anelli et~al\mbox{.}(2021)]%
        {anelli2021federank}
\bibfield{author}{\bibinfo{person}{Vito~Walter Anelli}, \bibinfo{person}{Yashar Deldjoo}, \bibinfo{person}{Tommaso Di~Noia}, \bibinfo{person}{Antonio Ferrara}, {and} \bibinfo{person}{Fedelucio Narducci}.} \bibinfo{year}{2021}\natexlab{}.
\newblock \showarticletitle{Federank: User controlled feedback with federated recommender systems}. In \bibinfo{booktitle}{\emph{ECIR}}. \bibinfo{pages}{32--47}.
\newblock


\bibitem[Bobadilla et~al\mbox{.}(2013)]%
        {bobadilla2013recommender}
\bibfield{author}{\bibinfo{person}{Jes{\'u}s Bobadilla}, \bibinfo{person}{Fernando Ortega}, \bibinfo{person}{Antonio Hernando}, {and} \bibinfo{person}{Abraham Guti{\'e}rrez}.} \bibinfo{year}{2013}\natexlab{}.
\newblock \showarticletitle{Recommender systems survey}.
\newblock \bibinfo{journal}{\emph{Knowledge-based systems}}  \bibinfo{volume}{46} (\bibinfo{year}{2013}), \bibinfo{pages}{109--132}.
\newblock


\bibitem[Chai et~al\mbox{.}(2020)]%
        {chai2020secure}
\bibfield{author}{\bibinfo{person}{Di Chai}, \bibinfo{person}{Leye Wang}, \bibinfo{person}{Kai Chen}, {and} \bibinfo{person}{Qiang Yang}.} \bibinfo{year}{2020}\natexlab{}.
\newblock \showarticletitle{Secure federated matrix factorization}.
\newblock \bibinfo{journal}{\emph{IEEE Intelligent Systems}} \bibinfo{volume}{36}, \bibinfo{number}{5} (\bibinfo{year}{2020}), \bibinfo{pages}{11--20}.
\newblock


\bibitem[Chen et~al\mbox{.}(2020)]%
        {chen2020simple}
\bibfield{author}{\bibinfo{person}{Ting Chen}, \bibinfo{person}{Simon Kornblith}, \bibinfo{person}{Mohammad Norouzi}, {and} \bibinfo{person}{Geoffrey Hinton}.} \bibinfo{year}{2020}\natexlab{}.
\newblock \showarticletitle{A simple framework for contrastive learning of visual representations}. In \bibinfo{booktitle}{\emph{International conference on machine learning}}. PMLR, \bibinfo{pages}{1597--1607}.
\newblock


\bibitem[Covington et~al\mbox{.}(2016)]%
        {covington2016deep}
\bibfield{author}{\bibinfo{person}{Paul Covington}, \bibinfo{person}{Jay Adams}, {and} \bibinfo{person}{Emre Sargin}.} \bibinfo{year}{2016}\natexlab{}.
\newblock \showarticletitle{Deep neural networks for youtube recommendations}. In \bibinfo{booktitle}{\emph{RecSys}}. \bibinfo{pages}{191--198}.
\newblock


\bibitem[Dwork et~al\mbox{.}(2006)]%
        {dwork2006calibrating}
\bibfield{author}{\bibinfo{person}{Cynthia Dwork}, \bibinfo{person}{Frank McSherry}, \bibinfo{person}{Kobbi Nissim}, {and} \bibinfo{person}{Adam Smith}.} \bibinfo{year}{2006}\natexlab{}.
\newblock \showarticletitle{Calibrating noise to sensitivity in private data analysis}. In \bibinfo{booktitle}{\emph{TCC}}. \bibinfo{pages}{265--284}.
\newblock


\bibitem[Glorot and Bengio(2010)]%
        {glorot2010understanding}
\bibfield{author}{\bibinfo{person}{Xavier Glorot} {and} \bibinfo{person}{Yoshua Bengio}.} \bibinfo{year}{2010}\natexlab{}.
\newblock \showarticletitle{Understanding the difficulty of training deep feedforward neural networks}. In \bibinfo{booktitle}{\emph{JMLR}}. \bibinfo{pages}{249--256}.
\newblock


\bibitem[Guo et~al\mbox{.}(2017)]%
        {guo2017deepfm}
\bibfield{author}{\bibinfo{person}{Huifeng Guo}, \bibinfo{person}{Ruiming Tang}, \bibinfo{person}{Yunming Ye}, \bibinfo{person}{Zhenguo Li}, {and} \bibinfo{person}{Xiuqiang He}.} \bibinfo{year}{2017}\natexlab{}.
\newblock \showarticletitle{DeepFM: a factorization-machine based neural network for CTR prediction}.
\newblock \bibinfo{journal}{\emph{arXiv preprint arXiv:1703.04247}} (\bibinfo{year}{2017}).
\newblock


\bibitem[Gutmann and Hyv{\"a}rinen(2010)]%
        {gutmann2010noise}
\bibfield{author}{\bibinfo{person}{Michael Gutmann} {and} \bibinfo{person}{Aapo Hyv{\"a}rinen}.} \bibinfo{year}{2010}\natexlab{}.
\newblock \showarticletitle{Noise-contrastive estimation: A new estimation principle for unnormalized statistical models}. In \bibinfo{booktitle}{\emph{AISTATS}}. \bibinfo{pages}{297--304}.
\newblock


\bibitem[He et~al\mbox{.}(2020)]%
        {he2020lightgcn}
\bibfield{author}{\bibinfo{person}{Xiangnan He}, \bibinfo{person}{Kuan Deng}, \bibinfo{person}{Xiang Wang}, \bibinfo{person}{Yan Li}, \bibinfo{person}{Yongdong Zhang}, {and} \bibinfo{person}{Meng Wang}.} \bibinfo{year}{2020}\natexlab{}.
\newblock \showarticletitle{Lightgcn: Simplifying and powering graph convolution network for recommendation}. In \bibinfo{booktitle}{\emph{SIGIR}}. \bibinfo{pages}{639--648}.
\newblock


\bibitem[He et~al\mbox{.}(2017)]%
        {10.1145/3038912.3052569}
\bibfield{author}{\bibinfo{person}{Xiangnan He}, \bibinfo{person}{Lizi Liao}, \bibinfo{person}{Hanwang Zhang}, \bibinfo{person}{Liqiang Nie}, \bibinfo{person}{Xia Hu}, {and} \bibinfo{person}{Tat-Seng Chua}.} \bibinfo{year}{2017}\natexlab{}.
\newblock \showarticletitle{Neural Collaborative Filtering}. In \bibinfo{booktitle}{\emph{WWW}}. \bibinfo{pages}{173–182}.
\newblock


\bibitem[Karimireddy et~al\mbox{.}(2019)]%
        {karimireddy2019scaffold}
\bibfield{author}{\bibinfo{person}{Sai~Praneeth Karimireddy}, \bibinfo{person}{Satyen Kale}, \bibinfo{person}{Mehryar Mohri}, \bibinfo{person}{Sashank~J Reddi}, \bibinfo{person}{Sebastian~U Stich}, {and} \bibinfo{person}{Ananda~Theertha Suresh}.} \bibinfo{year}{2019}\natexlab{}.
\newblock \showarticletitle{SCAFFOLD: Stochastic Controlled Averaging for On-Device Federated Learning.}
\newblock  (\bibinfo{year}{2019}).
\newblock


\bibitem[Kingma and Ba(2014)]%
        {kingma2014adam}
\bibfield{author}{\bibinfo{person}{Diederik~P Kingma} {and} \bibinfo{person}{Jimmy Ba}.} \bibinfo{year}{2014}\natexlab{}.
\newblock \showarticletitle{Adam: A method for stochastic optimization}.
\newblock \bibinfo{journal}{\emph{arXiv preprint arXiv:1412.6980}} (\bibinfo{year}{2014}).
\newblock


\bibitem[Koren et~al\mbox{.}(2009)]%
        {koren2009matrix}
\bibfield{author}{\bibinfo{person}{Yehuda Koren}, \bibinfo{person}{Robert Bell}, {and} \bibinfo{person}{Chris Volinsky}.} \bibinfo{year}{2009}\natexlab{}.
\newblock \showarticletitle{Matrix factorization techniques for recommender systems}.
\newblock \bibinfo{journal}{\emph{Computer}} \bibinfo{volume}{42}, \bibinfo{number}{8} (\bibinfo{year}{2009}), \bibinfo{pages}{30--37}.
\newblock


\bibitem[Li et~al\mbox{.}(2018)]%
        {li2018deeper}
\bibfield{author}{\bibinfo{person}{Qimai Li}, \bibinfo{person}{Zhichao Han}, {and} \bibinfo{person}{Xiao-Ming Wu}.} \bibinfo{year}{2018}\natexlab{}.
\newblock \showarticletitle{Deeper insights into graph convolutional networks for semi-supervised learning}. In \bibinfo{booktitle}{\emph{AAAI}}.
\newblock


\bibitem[Li et~al\mbox{.}(2019)]%
        {li2019convergence}
\bibfield{author}{\bibinfo{person}{Xiang Li}, \bibinfo{person}{Kaixuan Huang}, \bibinfo{person}{Wenhao Yang}, \bibinfo{person}{Shusen Wang}, {and} \bibinfo{person}{Zhihua Zhang}.} \bibinfo{year}{2019}\natexlab{}.
\newblock \showarticletitle{On the convergence of fedavg on non-iid data}.
\newblock \bibinfo{journal}{\emph{arXiv preprint arXiv:1907.02189}} (\bibinfo{year}{2019}).
\newblock


\bibitem[Liang et~al\mbox{.}(2016)]%
        {liang2016modeling}
\bibfield{author}{\bibinfo{person}{Dawen Liang}, \bibinfo{person}{Laurent Charlin}, \bibinfo{person}{James McInerney}, {and} \bibinfo{person}{David~M Blei}.} \bibinfo{year}{2016}\natexlab{}.
\newblock \showarticletitle{Modeling user exposure in recommendation}. In \bibinfo{booktitle}{\emph{WWW}}. \bibinfo{pages}{951--961}.
\newblock


\bibitem[Lin et~al\mbox{.}(2020)]%
        {lin2020fedrec}
\bibfield{author}{\bibinfo{person}{Guanyu Lin}, \bibinfo{person}{Feng Liang}, \bibinfo{person}{Weike Pan}, {and} \bibinfo{person}{Zhong Ming}.} \bibinfo{year}{2020}\natexlab{}.
\newblock \showarticletitle{Fedrec: Federated recommendation with explicit feedback}.
\newblock \bibinfo{journal}{\emph{IEEE Intelligent Systems}} \bibinfo{volume}{36}, \bibinfo{number}{5} (\bibinfo{year}{2020}), \bibinfo{pages}{21--30}.
\newblock


\bibitem[Liu et~al\mbox{.}(2022)]%
        {10.1145/3501815}
\bibfield{author}{\bibinfo{person}{Zhiwei Liu}, \bibinfo{person}{Liangwei Yang}, \bibinfo{person}{Ziwei Fan}, \bibinfo{person}{Hao Peng}, {and} \bibinfo{person}{Philip~S. Yu}.} \bibinfo{year}{2022}\natexlab{}.
\newblock \showarticletitle{Federated Social Recommendation with Graph Neural Network}.
\newblock \bibinfo{journal}{\emph{ACM Trans. Intell. Syst. Technol.}} \bibinfo{volume}{13}, \bibinfo{number}{4}, Article \bibinfo{articleno}{55} (\bibinfo{date}{aug} \bibinfo{year}{2022}), \bibinfo{numpages}{24}~pages.
\newblock
\showISSN{2157-6904}
\urldef\tempurl%
\url{https://doi.org/10.1145/3501815}
\showDOI{\tempurl}


\bibitem[Luo et~al\mbox{.}(2022)]%
        {10.1145/3511808.3557668}
\bibfield{author}{\bibinfo{person}{Sichun Luo}, \bibinfo{person}{Yuanzhang Xiao}, {and} \bibinfo{person}{Linqi Song}.} \bibinfo{year}{2022}\natexlab{}.
\newblock \showarticletitle{Personalized Federated Recommendation via Joint Representation Learning, User Clustering, and Model Adaptation}. In \bibinfo{booktitle}{\emph{CIKM}}. \bibinfo{pages}{4289–4293}.
\newblock


\bibitem[McMahan et~al\mbox{.}(2017)]%
        {mcmahan2017communication}
\bibfield{author}{\bibinfo{person}{Brendan McMahan}, \bibinfo{person}{Eider Moore}, \bibinfo{person}{Daniel Ramage}, \bibinfo{person}{Seth Hampson}, {and} \bibinfo{person}{Blaise~Aguera y Arcas}.} \bibinfo{year}{2017}\natexlab{}.
\newblock \showarticletitle{Communication-efficient learning of deep networks from decentralized data}. In \bibinfo{booktitle}{\emph{Artificial intelligence and statistics}}. PMLR, \bibinfo{pages}{1273--1282}.
\newblock


\bibitem[Qu et~al\mbox{.}(2023)]%
        {qu2023semi}
\bibfield{author}{\bibinfo{person}{Liang Qu}, \bibinfo{person}{Ningzhi Tang}, \bibinfo{person}{Ruiqi Zheng}, \bibinfo{person}{Quoc Viet~Hung Nguyen}, \bibinfo{person}{Zi Huang}, \bibinfo{person}{Yuhui Shi}, {and} \bibinfo{person}{Hongzhi Yin}.} \bibinfo{year}{2023}\natexlab{}.
\newblock \showarticletitle{Semi-decentralized Federated Ego Graph Learning for Recommendation}. In \bibinfo{booktitle}{\emph{WWW}}.
\newblock


\bibitem[Qu et~al\mbox{.}(2022)]%
        {qu2022single}
\bibfield{author}{\bibinfo{person}{Liang Qu}, \bibinfo{person}{Yonghong Ye}, \bibinfo{person}{Ningzhi Tang}, \bibinfo{person}{Lixin Zhang}, \bibinfo{person}{Yuhui Shi}, {and} \bibinfo{person}{Hongzhi Yin}.} \bibinfo{year}{2022}\natexlab{}.
\newblock \showarticletitle{Single-shot embedding dimension search in recommender system}. In \bibinfo{booktitle}{\emph{SIGIR}}. \bibinfo{pages}{513--522}.
\newblock


\bibitem[Ramlatchan et~al\mbox{.}(2018)]%
        {ramlatchan2018survey}
\bibfield{author}{\bibinfo{person}{Andy Ramlatchan}, \bibinfo{person}{Mengyun Yang}, \bibinfo{person}{Quan Liu}, \bibinfo{person}{Min Li}, \bibinfo{person}{Jianxin Wang}, {and} \bibinfo{person}{Yaohang Li}.} \bibinfo{year}{2018}\natexlab{}.
\newblock \showarticletitle{A survey of matrix completion methods for recommendation systems}.
\newblock \bibinfo{journal}{\emph{Big Data Mining and Analytics}} \bibinfo{volume}{1}, \bibinfo{number}{4} (\bibinfo{year}{2018}), \bibinfo{pages}{308--323}.
\newblock


\bibitem[Rendle et~al\mbox{.}(2012)]%
        {rendle2012bpr}
\bibfield{author}{\bibinfo{person}{Steffen Rendle}, \bibinfo{person}{Christoph Freudenthaler}, \bibinfo{person}{Zeno Gantner}, {and} \bibinfo{person}{Lars Schmidt-Thieme}.} \bibinfo{year}{2012}\natexlab{}.
\newblock \showarticletitle{BPR: Bayesian personalized ranking from implicit feedback}.
\newblock \bibinfo{journal}{\emph{arXiv preprint arXiv:1205.2618}} (\bibinfo{year}{2012}).
\newblock


\bibitem[Sun et~al\mbox{.}(2022)]%
        {sun2022survey}
\bibfield{author}{\bibinfo{person}{Zehua Sun}, \bibinfo{person}{Yonghui Xu}, \bibinfo{person}{Yong Liu}, \bibinfo{person}{Wei He}, \bibinfo{person}{Yali Jiang}, \bibinfo{person}{Fangzhao Wu}, {and} \bibinfo{person}{Lizhen Cui}.} \bibinfo{year}{2022}\natexlab{}.
\newblock \showarticletitle{A Survey on Federated Recommendation Systems}.
\newblock \bibinfo{journal}{\emph{arXiv preprint arXiv:2301.00767}} (\bibinfo{year}{2022}).
\newblock


\bibitem[Wang et~al\mbox{.}(2022)]%
        {wang2022fast}
\bibfield{author}{\bibinfo{person}{Qinyong Wang}, \bibinfo{person}{Hongzhi Yin}, \bibinfo{person}{Tong Chen}, \bibinfo{person}{Junliang Yu}, \bibinfo{person}{Alexander Zhou}, {and} \bibinfo{person}{Xiangliang Zhang}.} \bibinfo{year}{2022}\natexlab{}.
\newblock \showarticletitle{Fast-adapting and privacy-preserving federated recommender system}.
\newblock \bibinfo{journal}{\emph{The VLDB Journal}} \bibinfo{volume}{31}, \bibinfo{number}{5} (\bibinfo{year}{2022}), \bibinfo{pages}{877--896}.
\newblock


\bibitem[Wang et~al\mbox{.}(2019)]%
        {KGAT19}
\bibfield{author}{\bibinfo{person}{Xiang Wang}, \bibinfo{person}{Xiangnan He}, \bibinfo{person}{Yixin Cao}, \bibinfo{person}{Meng Liu}, {and} \bibinfo{person}{Tat-Seng Chua}.} \bibinfo{year}{2019}\natexlab{}.
\newblock \showarticletitle{KGAT: Knowledge Graph Attention Network for Recommendation}. In \bibinfo{booktitle}{\emph{{KDD}}}.
\newblock


\bibitem[Wu et~al\mbox{.}(2022b)]%
        {wu2022federated}
\bibfield{author}{\bibinfo{person}{Chuhan Wu}, \bibinfo{person}{Fangzhao Wu}, \bibinfo{person}{Lingjuan Lyu}, \bibinfo{person}{Tao Qi}, \bibinfo{person}{Yongfeng Huang}, {and} \bibinfo{person}{Xing Xie}.} \bibinfo{year}{2022}\natexlab{b}.
\newblock \showarticletitle{A federated graph neural network framework for privacy-preserving personalization}.
\newblock \bibinfo{journal}{\emph{Nature Communications}} \bibinfo{volume}{13}, \bibinfo{number}{1} (\bibinfo{year}{2022}), \bibinfo{pages}{3091}.
\newblock


\bibitem[Wu et~al\mbox{.}(2021)]%
        {wu2021self}
\bibfield{author}{\bibinfo{person}{Jiancan Wu}, \bibinfo{person}{Xiang Wang}, \bibinfo{person}{Fuli Feng}, \bibinfo{person}{Xiangnan He}, \bibinfo{person}{Liang Chen}, \bibinfo{person}{Jianxun Lian}, {and} \bibinfo{person}{Xing Xie}.} \bibinfo{year}{2021}\natexlab{}.
\newblock \showarticletitle{Self-supervised graph learning for recommendation}. In \bibinfo{booktitle}{\emph{SIGIR}}. \bibinfo{pages}{726--735}.
\newblock


\bibitem[Wu et~al\mbox{.}(2022a)]%
        {wu2022graph}
\bibfield{author}{\bibinfo{person}{Shiwen Wu}, \bibinfo{person}{Fei Sun}, \bibinfo{person}{Wentao Zhang}, \bibinfo{person}{Xu Xie}, {and} \bibinfo{person}{Bin Cui}.} \bibinfo{year}{2022}\natexlab{a}.
\newblock \showarticletitle{Graph neural networks in recommender systems: a survey}.
\newblock \bibinfo{journal}{\emph{Comput. Surveys}} \bibinfo{volume}{55}, \bibinfo{number}{5} (\bibinfo{year}{2022}), \bibinfo{pages}{1--37}.
\newblock


\bibitem[Yang et~al\mbox{.}(2020)]%
        {yang2020federated}
\bibfield{author}{\bibinfo{person}{Liu Yang}, \bibinfo{person}{Ben Tan}, \bibinfo{person}{Vincent~W Zheng}, \bibinfo{person}{Kai Chen}, {and} \bibinfo{person}{Qiang Yang}.} \bibinfo{year}{2020}\natexlab{}.
\newblock \showarticletitle{Federated recommendation systems}.
\newblock In \bibinfo{booktitle}{\emph{Federated Learning}}. \bibinfo{publisher}{Springer}, \bibinfo{pages}{225--239}.
\newblock


\bibitem[Yang et~al\mbox{.}(2019)]%
        {yang2019federated}
\bibfield{author}{\bibinfo{person}{Qiang Yang}, \bibinfo{person}{Yang Liu}, \bibinfo{person}{Yong Cheng}, \bibinfo{person}{Yan Kang}, \bibinfo{person}{Tianjian Chen}, {and} \bibinfo{person}{Han Yu}.} \bibinfo{year}{2019}\natexlab{}.
\newblock \showarticletitle{Federated learning}.
\newblock \bibinfo{journal}{\emph{Synthesis Lectures on Artificial Intelligence and Machine Learning}} \bibinfo{volume}{13}, \bibinfo{number}{3} (\bibinfo{year}{2019}), \bibinfo{pages}{1--207}.
\newblock


\bibitem[Yin et~al\mbox{.}(2015)]%
        {yin2015joint}
\bibfield{author}{\bibinfo{person}{Hongzhi Yin}, \bibinfo{person}{Bin Cui}, \bibinfo{person}{Zi Huang}, \bibinfo{person}{Weiqing Wang}, \bibinfo{person}{Xian Wu}, {and} \bibinfo{person}{Xiaofang Zhou}.} \bibinfo{year}{2015}\natexlab{}.
\newblock \showarticletitle{Joint modeling of users' interests and mobility patterns for point-of-interest recommendation}. In \bibinfo{booktitle}{\emph{MM}}. \bibinfo{pages}{819--822}.
\newblock


\bibitem[Yin et~al\mbox{.}(2024)]%
        {yin2024device}
\bibfield{author}{\bibinfo{person}{Hongzhi Yin}, \bibinfo{person}{Liang Qu}, \bibinfo{person}{Tong Chen}, \bibinfo{person}{Wei Yuan}, \bibinfo{person}{Ruiqi Zheng}, \bibinfo{person}{Jing Long}, \bibinfo{person}{Xin Xia}, \bibinfo{person}{Yuhui Shi}, {and} \bibinfo{person}{Chengqi Zhang}.} \bibinfo{year}{2024}\natexlab{}.
\newblock \showarticletitle{On-Device Recommender Systems: A Comprehensive Survey}.
\newblock \bibinfo{journal}{\emph{arXiv preprint arXiv:2401.11441}} (\bibinfo{year}{2024}).
\newblock


\bibitem[Ying et~al\mbox{.}(2018)]%
        {ying2018graph}
\bibfield{author}{\bibinfo{person}{Rex Ying}, \bibinfo{person}{Ruining He}, \bibinfo{person}{Kaifeng Chen}, \bibinfo{person}{Pong Eksombatchai}, \bibinfo{person}{William~L Hamilton}, {and} \bibinfo{person}{Jure Leskovec}.} \bibinfo{year}{2018}\natexlab{}.
\newblock \showarticletitle{Graph convolutional neural networks for web-scale recommender systems}. In \bibinfo{booktitle}{\emph{KDD}}. \bibinfo{pages}{974--983}.
\newblock


\bibitem[Yuan et~al\mbox{.}(2023a)]%
        {yuan2023manipulating}
\bibfield{author}{\bibinfo{person}{Wei Yuan}, \bibinfo{person}{Quoc Viet~Hung Nguyen}, \bibinfo{person}{Tieke He}, \bibinfo{person}{Liang Chen}, {and} \bibinfo{person}{Hongzhi Yin}.} \bibinfo{year}{2023}\natexlab{a}.
\newblock \showarticletitle{Manipulating Federated Recommender Systems: Poisoning with Synthetic Users and Its Countermeasures}.
\newblock \bibinfo{journal}{\emph{arXiv preprint arXiv:2304.03054}} (\bibinfo{year}{2023}).
\newblock


\bibitem[Yuan et~al\mbox{.}(2023b)]%
        {yuan2023hetefedrec}
\bibfield{author}{\bibinfo{person}{Wei Yuan}, \bibinfo{person}{Liang Qu}, \bibinfo{person}{Lizhen Cui}, \bibinfo{person}{Yongxin Tong}, \bibinfo{person}{Xiaofang Zhou}, {and} \bibinfo{person}{Hongzhi Yin}.} \bibinfo{year}{2023}\natexlab{b}.
\newblock \showarticletitle{HeteFedRec: Federated Recommender Systems with Model Heterogeneity}.
\newblock \bibinfo{journal}{\emph{arXiv preprint arXiv:2307.12810}} (\bibinfo{year}{2023}).
\newblock


\bibitem[Yuan et~al\mbox{.}(2023c)]%
        {yuan2023interaction}
\bibfield{author}{\bibinfo{person}{Wei Yuan}, \bibinfo{person}{Chaoqun Yang}, \bibinfo{person}{Quoc Viet~Hung Nguyen}, \bibinfo{person}{Lizhen Cui}, \bibinfo{person}{Tieke He}, {and} \bibinfo{person}{Hongzhi Yin}.} \bibinfo{year}{2023}\natexlab{c}.
\newblock \showarticletitle{Interaction-level membership inference attack against federated recommender systems}.
\newblock \bibinfo{journal}{\emph{arXiv preprint arXiv:2301.10964}} (\bibinfo{year}{2023}).
\newblock


\bibitem[Yuan et~al\mbox{.}(2023d)]%
        {yuan2023federated}
\bibfield{author}{\bibinfo{person}{Wei Yuan}, \bibinfo{person}{Hongzhi Yin}, \bibinfo{person}{Fangzhao Wu}, \bibinfo{person}{Shijie Zhang}, \bibinfo{person}{Tieke He}, {and} \bibinfo{person}{Hao Wang}.} \bibinfo{year}{2023}\natexlab{d}.
\newblock \showarticletitle{Federated unlearning for on-device recommendation}. In \bibinfo{booktitle}{\emph{WWW}}. \bibinfo{pages}{393--401}.
\newblock


\bibitem[Zhang et~al\mbox{.}(2021a)]%
        {zhang2021subgraph}
\bibfield{author}{\bibinfo{person}{Ke Zhang}, \bibinfo{person}{Carl Yang}, \bibinfo{person}{Xiaoxiao Li}, \bibinfo{person}{Lichao Sun}, {and} \bibinfo{person}{Siu~Ming Yiu}.} \bibinfo{year}{2021}\natexlab{a}.
\newblock \showarticletitle{Subgraph federated learning with missing neighbor generation}.
\newblock \bibinfo{journal}{\emph{Advances in Neural Information Processing Systems}}  \bibinfo{volume}{34} (\bibinfo{year}{2021}), \bibinfo{pages}{6671--6682}.
\newblock


\bibitem[Zhang et~al\mbox{.}(2019)]%
        {zhang2019deep}
\bibfield{author}{\bibinfo{person}{Shuai Zhang}, \bibinfo{person}{Lina Yao}, \bibinfo{person}{Aixin Sun}, {and} \bibinfo{person}{Yi Tay}.} \bibinfo{year}{2019}\natexlab{}.
\newblock \showarticletitle{Deep learning based recommender system: A survey and new perspectives}.
\newblock \bibinfo{journal}{\emph{ACM Computing Surveys (CSUR)}} \bibinfo{volume}{52}, \bibinfo{number}{1} (\bibinfo{year}{2019}), \bibinfo{pages}{1--38}.
\newblock


\bibitem[Zhang et~al\mbox{.}(2021b)]%
        {zhang2021graph}
\bibfield{author}{\bibinfo{person}{Shijie Zhang}, \bibinfo{person}{Hongzhi Yin}, \bibinfo{person}{Tong Chen}, \bibinfo{person}{Zi Huang}, \bibinfo{person}{Lizhen Cui}, {and} \bibinfo{person}{Xiangliang Zhang}.} \bibinfo{year}{2021}\natexlab{b}.
\newblock \showarticletitle{Graph embedding for recommendation against attribute inference attacks}. In \bibinfo{booktitle}{\emph{Proceedings of the Web Conference 2021}}. \bibinfo{pages}{3002--3014}.
\newblock


\bibitem[Zhang et~al\mbox{.}(2023)]%
        {zhang2023comprehensive}
\bibfield{author}{\bibinfo{person}{Shijie Zhang}, \bibinfo{person}{Wei Yuan}, {and} \bibinfo{person}{Hongzhi Yin}.} \bibinfo{year}{2023}\natexlab{}.
\newblock \showarticletitle{Comprehensive privacy analysis on federated recommender system against attribute inference attacks}.
\newblock \bibinfo{journal}{\emph{IEEE Transactions on Knowledge and Data Engineering}} (\bibinfo{year}{2023}).
\newblock


\bibitem[Zheng et~al\mbox{.}(2023)]%
        {zheng2023automl}
\bibfield{author}{\bibinfo{person}{Ruiqi Zheng}, \bibinfo{person}{Liang Qu}, \bibinfo{person}{Bin Cui}, \bibinfo{person}{Yuhui Shi}, {and} \bibinfo{person}{Hongzhi Yin}.} \bibinfo{year}{2023}\natexlab{}.
\newblock \showarticletitle{Automl for deep recommender systems: A survey}.
\newblock \bibinfo{journal}{\emph{ACM Transactions on Information Systems}} \bibinfo{volume}{41}, \bibinfo{number}{4} (\bibinfo{year}{2023}), \bibinfo{pages}{1--38}.
\newblock


\end{thebibliography}

\end{document}